\documentclass{kapproc} 

\usepackage{procps} 

\usepackage[dvips]{graphicx}

\upperandlowercase

\kluwerbib 

\begin{document}

\input BoxedEPS.tex
\SetEPSFDirectory{./}
\SetRokickiEPSFSpecial
\HideDisplacementBoxes

\articletitle[The local sub-mm luminosity functions]
{The local sub-mm luminosity functions and predictions 
from ASTRO-F/SIRTF to Herschel}

\author{
Stephen Serjeant$^1$, Diana Harrison$^2$
}

\affil{
$^1$School of Physical Sciences, University of Kent, Canterbury, Kent,
CT2 7NR\\ 
$^2$Dept. of Physics, Cavendish Laboratory, Madingley Road, Cambridge
CB3 0HE, UK
}
\email{}


\anxx{Serjeant, S., \& Harrison, D.}

\begin{abstract}
We present new
determinations of the local sub-mm luminosity functions. 
We find 
the local sub-mm luminosity density converging to 
$7.3\pm0.2\times10^{19}$ $h_{65}^{-1}$ W Hz$^{-1}$ Mpc$^{-3}$ at
$850\mu$m solving the ``sub-mm Olbers' Paradox.''
Using the sub-mm colour temperature relations from the SCUBA Local
Universe
Galaxy Survey, and the discovery of excess $450\mu$m excess emission
in these galaxies, we interpolate and extrapolate the IRAS detections
to make predictions of the SEDs of all 15411 PSC-z galaxies 
from $50-3000\mu$m. Despite the long
extrapolations we find excellent agreement with (a) the $90\mu$m
luminosity function of Serjeant et al. (2001), (b) the $850\mu$m
luminosity function of Dunne et al. (2000), (c) the mm-wave photometry
of Andreani \& Franceschini (1996); (d) the asymptotic differential
and integral source count predictions at $50-3000\mu$m by
Rowan-Robinson (2001). Remarkably, the local luminosity density and
the extragalactic background light together strongly constrain the
cosmic star formation history for a wide class of evolutionary
assumptions. We find that the extragalactic background light, the
$850\mu$m $8$mJy source counts, and the $\Omega_\ast$ constraints all
independently point to a decline in the comoving star formation rate
at $z>1$. 
\end{abstract}


\section{Introduction}
The evolution of the sub-mm galaxy population can be strongly
constrained 
by the integrated extragalactic background light, the local
multiwavelength luminosity functions, and the source counts. 
The local $850\mu$m luminosity
function was derived in  
the SCUBA Local Universe Galaxy Survey (SLUGS, Dunne et al. 2000) from
their SCUBA photometry of the IRAS Bright Galaxy Survey. A curious
aspect of their luminosity function was that the faint end slope was
not sufficiently shallow for the local luminosity density to converge,
which the authors referred to as the ``sub-mm Olbers' paradox''. This
is a pity from the point of view of modelling the high redshift
population, since the integrated extragalactic background light is a
key constraint. 
In order
to find the expected flattening of the luminosity function slope at
lower luminosities, the SLUGS 
survey is currently being extended with SCUBA photometry of
optically-selected galaxies. Meanwhile, several authors have
attempted to use the colour temperature -- luminosity relation found
in SLUGS to transform the $60\mu$m luminosity function to other
wavelengths, and hence constrain the high-redshift evolution
(e.g. Lagache, Dole \& Puget 2003, Chapman
et al. 2003). The 
discovery of an
additional excess component at $450\mu$m (Dunne \& Eales 2001)
relative to their initial colour temperature -- luminosity relation
has not so far been included in such models. 

\section{Models of PSC-z galaxies}
In this paper we take an
alternative approach to determining the multiwavelength local
luminosity functions. 
We model the spectral energy distributions (SEDs) of all 15411 
PSC-z galaxies (Saunders et al. 2000), constrained by all available
far-infrared and sub-mm colour-colour relations from SLUGS and
elsewhere.
This guarantees the correct local population mix at every wavelength
and minimises the assumptions about the trends of SED shape with
luminosity, and is sufficient to determine 
the $450\mu$m and $850\mu$m luminosities to better than 
a factor of $2$ in individual galaxies. 
This predicted photometry is more than sufficient 
to determine local luminosity functions.
For interpolations between these bands 
(e.g. predicted photometry from SIRTF, ASTRO-F, or Herschel) 
we express the SED as the sum
of two $\beta=2$ grey bodies, which the measured and/or predicted 
bands are just sufficient
to determine uniquely. 
More details can be found in
Serjeant \& Harrision (2003).

\begin{figure}[ht]
\vspace*{-1cm}
  \ForceWidth{3.0in}
  \hSlide{3.6cm}
  \BoxedEPSF{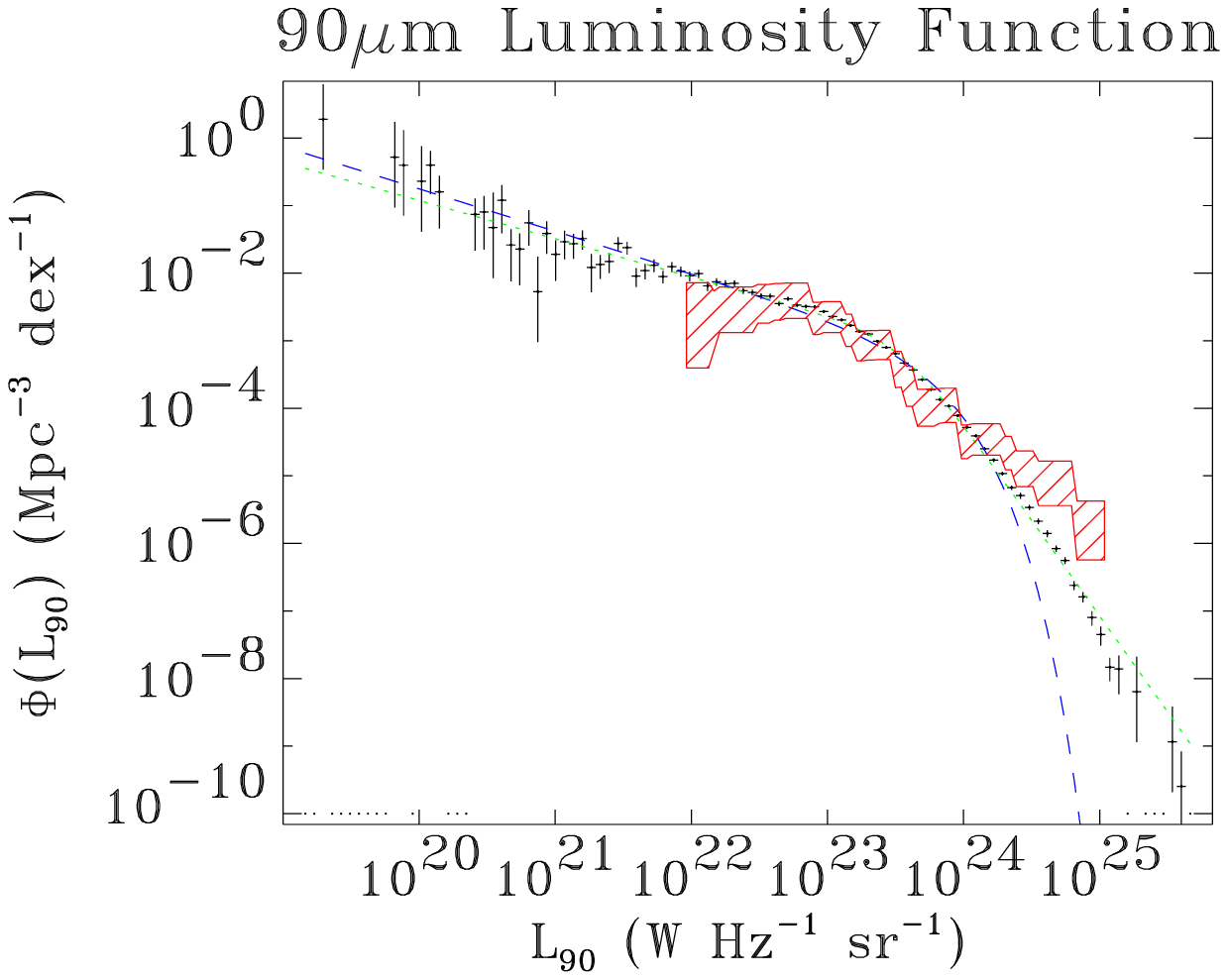}
\vspace*{-5.51cm}
  \ForceWidth{3.0in}
  \hSlide{-0.8cm}
  \BoxedEPSF{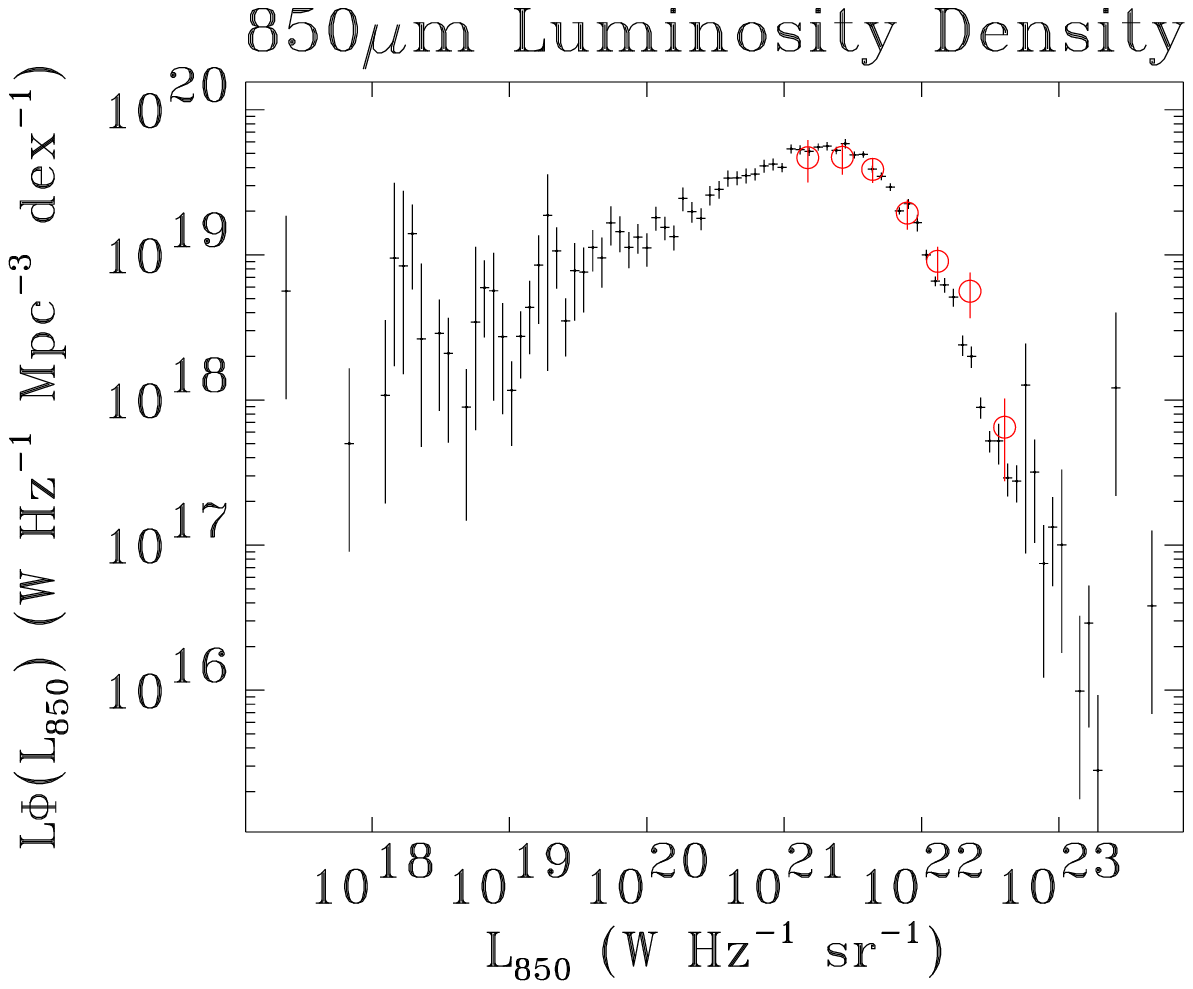}
\caption{
Left: projected local $850\mu$m luminosity density from PSC-Z (error
bars) assuming pure luminosity
evolution of $(1+z)^3$, compared with the directly measurements
from Dunne et al. (2000). 
Right: projected $90\mu$m luminosity function, compared
to the direct determination of Serjeant et al. 2001 (shaded area). 
}
\end{figure}


\section{Results}
There are few local galaxies with abundant multi-wavelength data to
test our models, but our models are consistent with the observed
$1.25$mm photometry of Andreani \& Franceschini (1996), the $175\mu$m
ISO Serendipity Survey (e.g. Stickel et al. 2000, also demonstrating
that the cool dust excess reported by Stickel et al. 2000 is
identical to that reported  by Dunne \& Eales 2001), and the SLUGS
galaxies with multi-wavelength data. The projected bright-end source
counts are also in excellent agreement with source count models
(Rowan-Robinson 2001).

\begin{figure}[ht]
\vspace*{-1cm}
  \ForceWidth{4.0in}
  \hSlide{-3cm}
  \BoxedEPSF{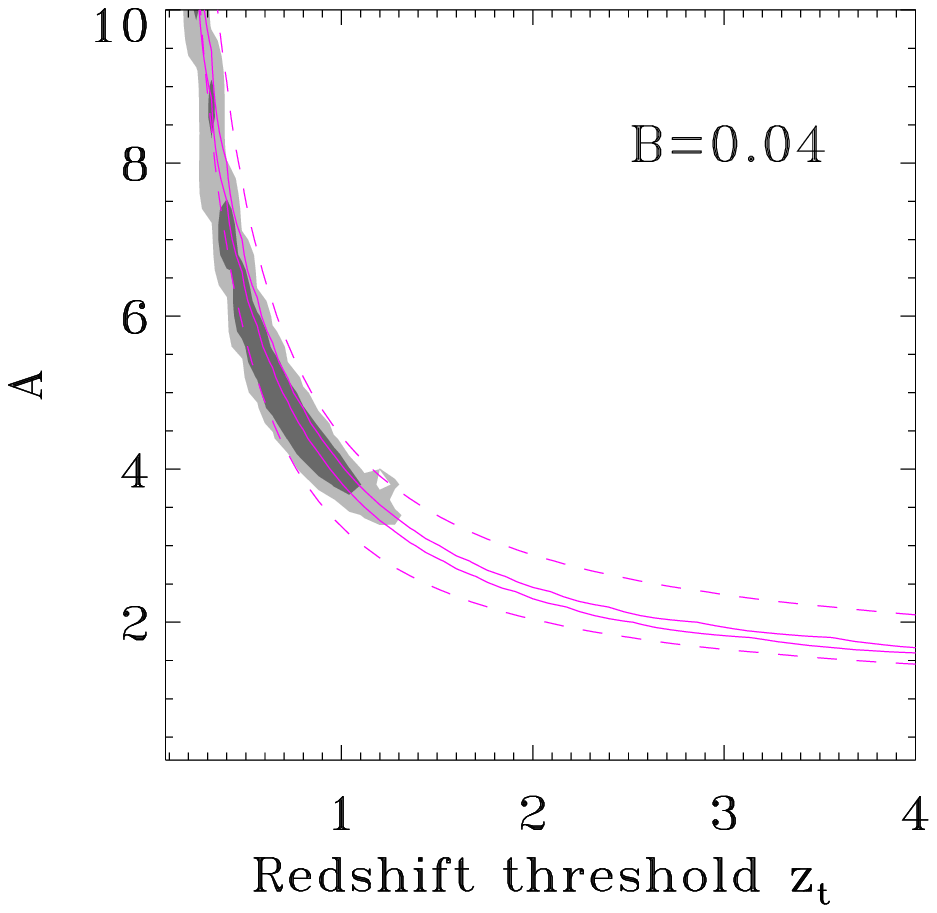}
\vspace*{-7.27cm}
  \ForceWidth{4.0in}
  \hSlide{4cm}
  \BoxedEPSF{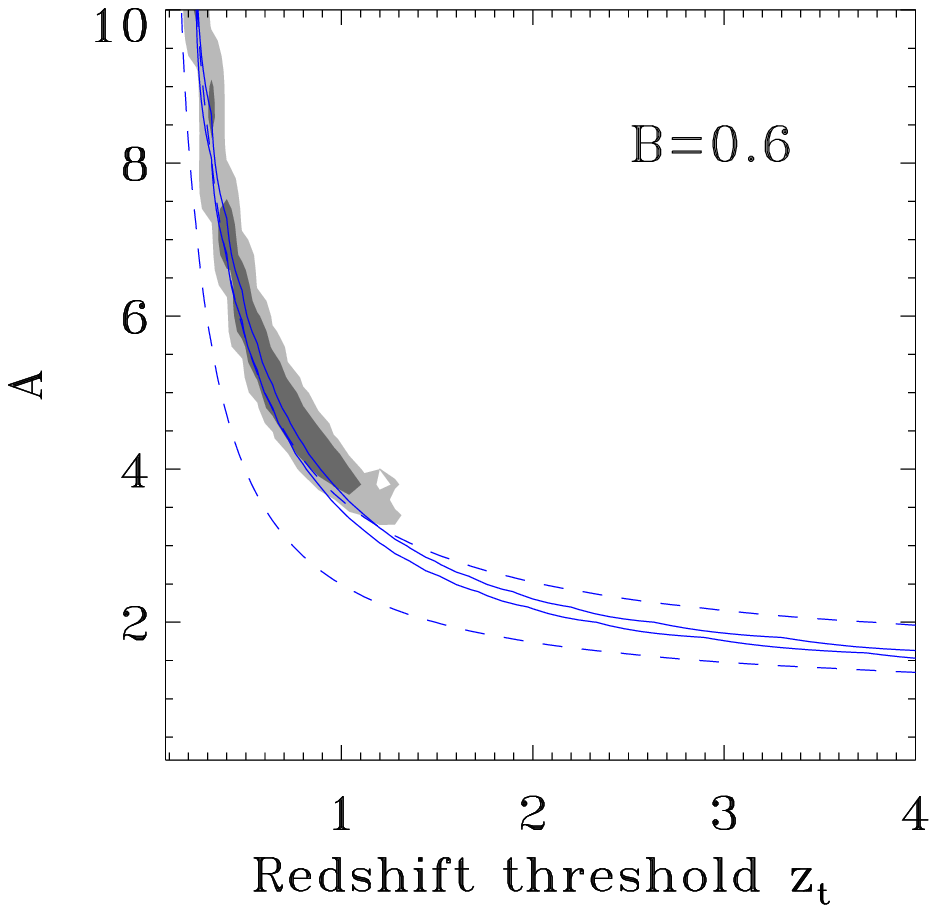}
\caption{Constraints on the parameterisation of the
$z\stackrel{<}{_\sim}2$ 
cosmic star formation history. The definitions of the parameters $A$
and $z_t$ are 
given in the text.
The shaded regions show the $68\%$ and
$95\%$ confidence regions (inner and outer region respectively) for
$A$ and $z_t$, marginalised over $B$. These constraints are derived
from only the extragalactic background light and our 
determination of the local spectral luminosity density. The dashed
lines show the {\em independent} constraints from
$\Omega_\ast=0.003\pm0.0009 h^{-1}$ (e.g. Lanzetta, Yahil, \& Fernandez-Soto
1996), and the full line shows the predicted 
$8$mJy $850\mu$m source count constraint of $N(>S)=320_{-100}^{+80}$
deg$^{-2}$ (Scott et al. 2002) assuming pure luminosity evolution. 
}
\end{figure}

\section{Discussion}
The PSC-z $850\mu$m luminosity function is in excellent agreement with
the direct determination of Dunne et al. (2000) and clearly shows the
convergence in the luminosity density (fig. 1). Other interpolated
luminosity functions can also be generated (fig. 1). The extragalactic
background light depends only on these local luminosity densities and
the cosmic star formation history, which we parameterise as
$\dot{\rho}(z)/\dot{\rho}(z=0)=(1+z)^A$ at $z<z_t$, and  $(1+z_t)^A
B^{z-z_t}$ at $z>z_t$. Not only do we obtain strong constraints on the
parameters $A,B,z_t$ (figs. 2 and 3) requiring a decline in comoving
star formation rate at $z>1$, but the $850\mu$m survey source
counts and $\Omega_\ast$ constraints both point to a similar
constraint. To reconcile this with other constraints on the cosmic
star formation history requires differential evolution in the
sub-mm population and/or a top-heavy IMF at high-$z$. 

\begin{figure}[ht]
  \ForceWidth{3in}
  \hSlide{-2.8cm}
  \BoxedEPSF{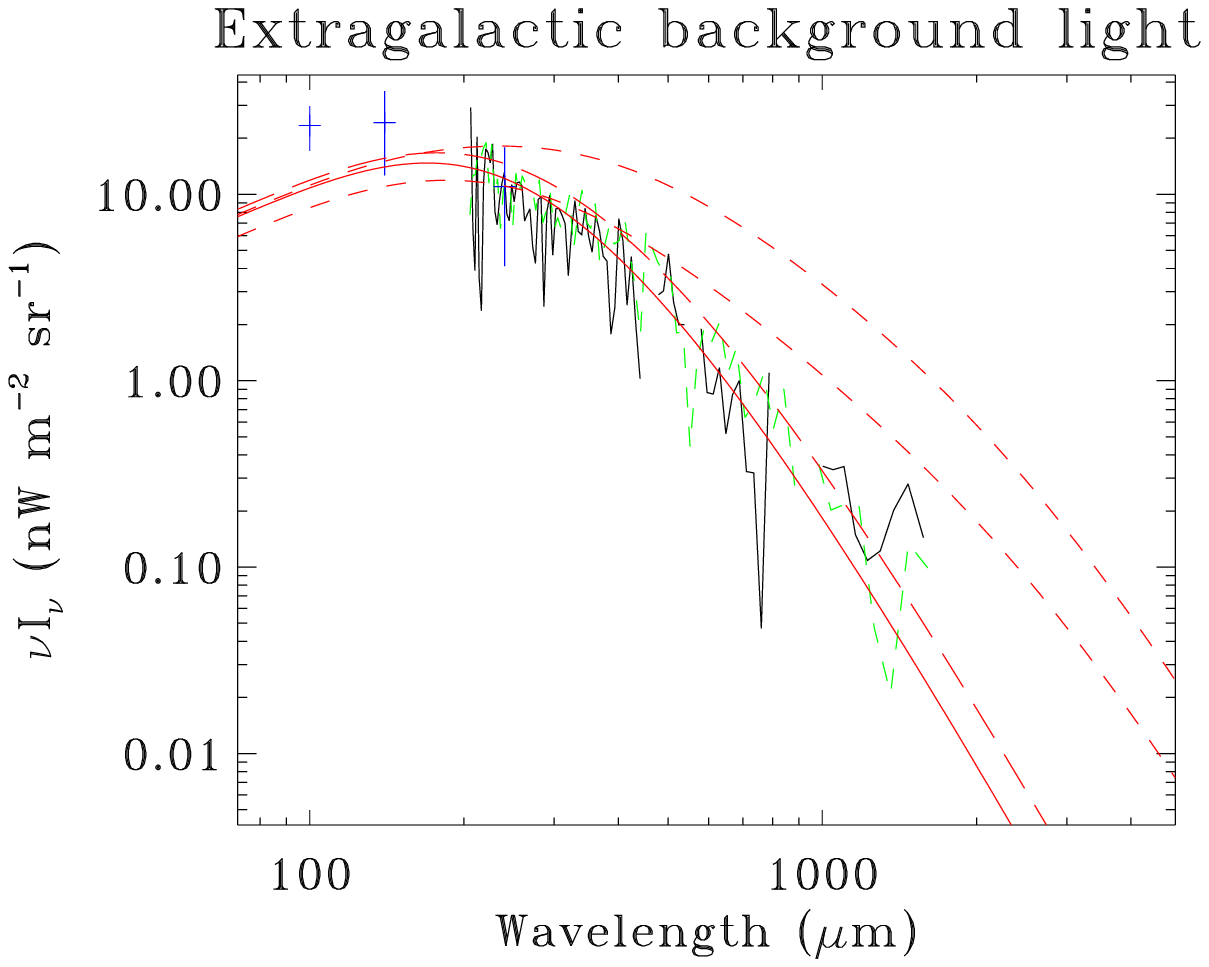}
\vspace*{-5.51cm}
  \ForceWidth{3in}
  \hSlide{5.2cm}
  \BoxedEPSF{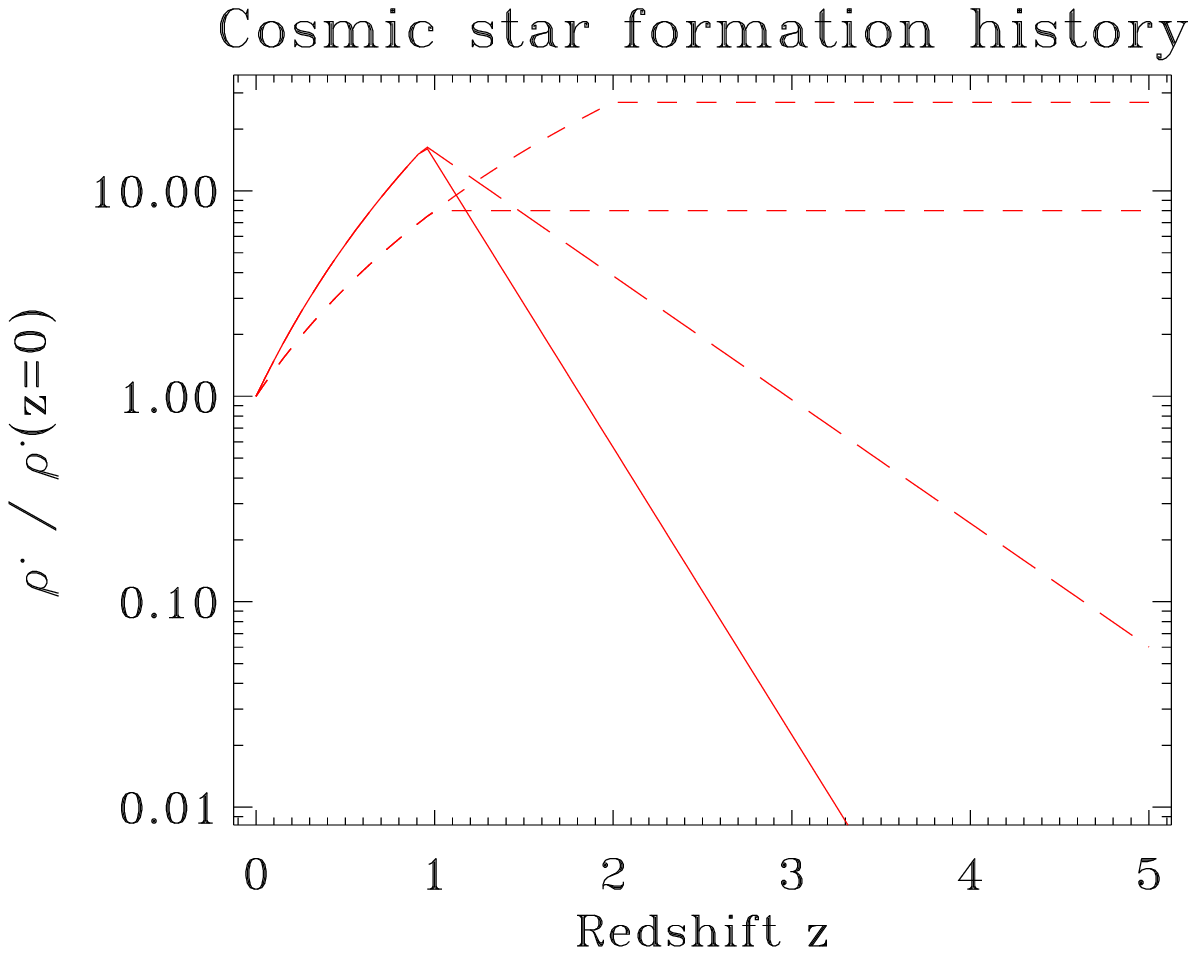}
\caption{
Left figure shows the 
extragalactic background light (data from Lagache et al. 1999)
modelled by the cosmic star formation history described in the text. 
The data longward of $200\mu$m plotted as broken
lines is the FIRAS spectrum: full line is
whole sky, and dashed line is Lockman Hole only. Also plotted are the
DIRBE data points, not included in our fitting. The smooth curves
are models of this spectrum, corresponding to cosmic star
formation histories plotted in the right hand figure. 
The full line is 
the global maximum likelihood fit,  and the long-dashed line has the
same 
parameters except an enhanced high-$z$ star formation rate ($B=0.25$)
which is marginally inconsistent with the
extragalactic background. The two short-dashed lines demonstrate 
selected alternative models: 
$z_t=1$ and $z_t=2$, both with $A=3$ and $B=1$. 
Note that in general the models with the
higher star formation rates at $z>2$ are also the models predicting the
larger background at wavelengths $\lambda\sim1$mm (including in
particular a comparison of the two short-dashed curves). In general,
models with high volume-averaged star formation rates at $z>2$
overpredict the sub-mm/mm-wave background. 
}
\end{figure}

\end{document}